\documentclass[12pt]{article}
\usepackage{amssymb}

\usepackage{graphicx}
\usepackage{epstopdf}
\usepackage{sw20aip}

\input{tcilatex}
\begin{document}

\date{12-14-2006}
\title{Coulomb oscillation in the hydrogen atom and molecule ion }
\author{Manfred Bucher \\
Physics Department, California State University, Fresno\\
Fresno, Califonia 93740-8031}
\maketitle

\begin{abstract}
Semiclassical oscillation of the electron through the nucleus of the $H$
atom yields both the exact energy and the correct orbital angular momentum
for $l=0$ quantum states. Similarly, electron oscillation through the nuclei
of $H_{2}^{+}$ accounts for a stable molecule ion with energy close to the
quantum mechanical solution. The small discrepancy arises from the neglect
of the electron's wave nature.

PACS numbers: 03.65.Sq, 31.10.+z, 31.20.Pv
\end{abstract}

\section{INTRODUCTION}

Two of the reasons why the old quantum theory of Bohr and Sommerfeld was
abandoned in the mid 1920s were the theory's failure to give the correct
multiplet structure of the hydrogen atom and the stability of the hydrogen
molecule ion, $H_{2}^{+}$.\cite{1} The old vector model of angular momentum%
\cite{2} gave, for a given principal quantum number $n$, sublevels with
angular quantum numbers $l=1,2,...,n$.\cite{3} Spectroscopic evidence,
however, showed multiplicities of spectral line splitting in a magnetic
field according to $l=0,1,...,n-1$. Max Jammer, in his review,\cite{2} notes
that ``the old quantum theory could never resolve this inconsistency.''

A treatment of the hydrogen molecule ion with Sommerfeld's quantization
conditions had been Wolfgang Pauli's doctoral thesis of 1922.\cite{4} Pauli
found its molecular binding energy to be positive (non-binding)---contrary
to (later) experimental findings. Martin Gutzwiller\cite{5} thinks that
``the solution of this problem can be rated, with only slight exaggeration,
as the most important in quantum mechanics, because if an energy level with
a [more] negative value [than of a free hydrogen atom] can be found, then
the chemical bond between two protons by a single electron has been
explained.''

Both dilemmas of the old quantum theory can be resolved, though, with a
single extension: an oscillation of the electron through the nucleus
(nuclei) of the atom (molecule). In essence this solution is already
formally included in Sommerfeld's theory of the hydrogen atom\cite{6} but
was explicitly omitted by Sommerfeld and his school as being \textit{\
unphysical}. The case in point, obtained with Sommerfeld's quantization
conditions for radial and angular motion, is a quantum state with zero
angular action, characterized by an angular quantum number $l$ = 0. What is
its orbit?

The geometry of an $nl$ Sommerfeld ellipse is given by its semimajor axis, $%
a_{nl}=(r_{B}/Z)n^{2}$, and semiminor axis, $b_{nl}=(r_{B}/Z)n\sqrt{l(l+1)}$.%
\cite{7} Here $r_{B}=h^{2}/4\pi ^{2}me^{2}$ is the Bohr radius in terms of
fundamental constants, serving as an atomic distance unit, and $Ze$ is the
nuclear charge. An $(n,0)$ orbit is thus a line ellipse with its nuclear
focus at one end and its empty focus at the other. This case was regarded as
unphysical because of the electron's collision with the nucleus---an
uncritical adaptation from celestrial mechanics. A closer inspection
confirms that a line ellipse with \textit{terminal} nuclear focus is indeed
unphysical---but for quite a different reason!\bigskip

\includegraphics[width=5in]{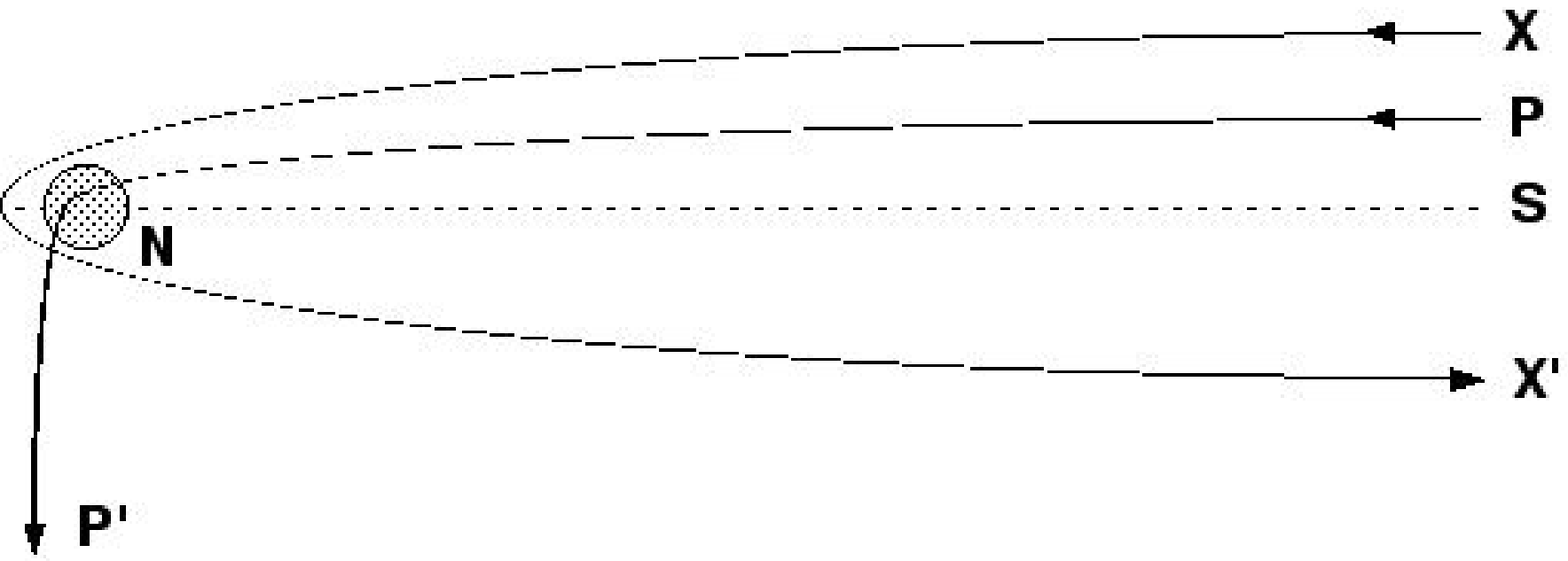}

\begin{quote}
Fig. 1. Partial trajectory of an extranuclear orbit $XX^{\prime }$\ and of a
penetrating orbit $PP^{\prime }$\ through nucleus $N$. The dotted line $S$\
shows the major symmetry axis of $XX^{\prime }$.\bigskip
\end{quote}

Leaving quantization conditions momentarily aside, what would happen if we 
\textit{continuously} decrease an angular quantum number $\lambda $ while
keeping the principal quantum number $n$ constant? We then would get more
and more slender ellipses with the same length of major axis, $2a_{n}$. By
basic electric theory, the nuclear Coulomb potential outside a \textit{\
finite-size} nucleus $N$ is given by the point potential as if all nuclear
charge, $+Ze$, was concentrated at the center of the nucleus. This holds as
long as the electron orbit stays outside the nucleus. Two borderline cases
are illustrated in Fig. 1. For a very small $\lambda $ value, say $0<\lambda
_{X}\ll 1$, we obtain a very slender elliptical orbit with partial
trajectory $XX^{\prime }$ about the nucleus. Further decrease of $\lambda $
to $\lambda _{P}$ \TEXTsymbol{<} $\lambda _{X}$ causes an intrusion of the
electron into the finite nucleus of radius $r_{N}$ (see trajectory $PN$ in
Fig. 1). Once inside, at a distance $r<r_{N}$ from the center, then, by
Gauss's law, only a fraction of the nuclear charge, $Z^{\prime }(r)e<Ze$,
acts on the electron via centripetal force.\cite{8} Accordingly, the
electron's exit trajectory $NP^{\prime }$ is no longer symmetric to its
approach trajectory $PN$ with respect to the major axis $S$ of the (partial)
ellipse $XX^{\prime }$. In the extreme case of a head-on penetration of the
nucleus, $\lambda =0$, there is no centripetal force at all! The electron
will then, with almost constant speed, traverse the nucleus, continue, with
decreasing speed, to the opposite turning point of its line orbit and revert
its motion periodically. We want to call the electron's straight-line
oscillation in the Coulomb potential of a finite-size nucleus a \textit{\
``Coulomb oscillator.''\bigskip }

\includegraphics[width=5in]{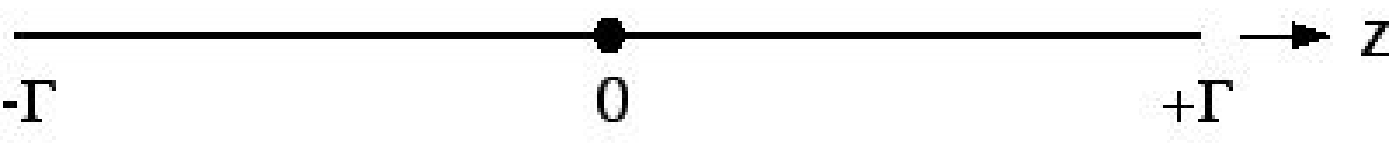}

\begin{quote}
Fig. 2. Line orbit of a Coulomb oscillator with nucleus at origin $0$ and
turning points at $\pm \Gamma $.\textit{\bigskip }
\end{quote}

\section{HYDROGEN ATOM}

For the formal treatment of the non-relativistic Coulomb oscillator we
designate the $z$ axis along the line orbit, with turning points at $z=\pm
\Gamma $ and nuclear position at $z=0$ (see Fig. 2). The electron's total
energy $E$ at position $z$ must equal the potential energy at a turning
point,

\begin{equation}
E=\frac{1}{2}mv^{2}-\frac{Ze^{2}}{|z|}=-\frac{Ze^{2}}{\Gamma }.  \tag{1}
\end{equation}

\noindent This gives the electron's speed along the $z$ axis,

\begin{equation}
v=\pm e\sqrt{\frac{2Z}{m}}\sqrt{\frac{1}{|z|}-\frac{1}{\Gamma }}.  \tag{2}
\end{equation}

\includegraphics[width=5in]{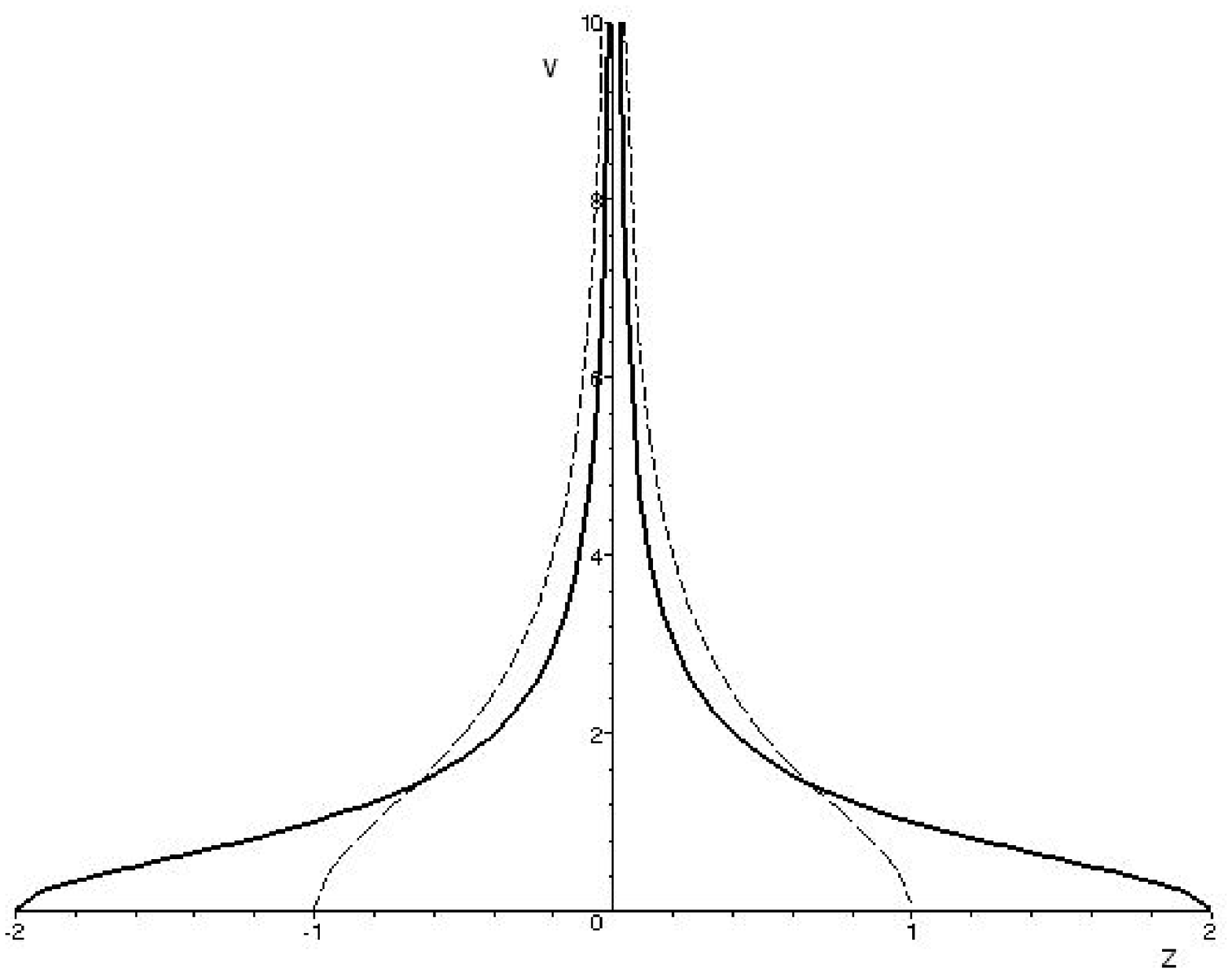}

\begin{quote}
Fig. 3. Axial electron speed $v$ vs. position $z$ of an electron in Coulomb
oscillation for the ground state, $n=1$, of a hydrogen atom $H$ ($Z=1$,
solid curve) and a helium ion $He^{+}$ ($Z=2$, dashed curve). The area under
one wing of each curve represents the radial action, $A=1h$.\bigskip
\end{quote}

\noindent Figure 3 displays its dependence on the axial position as a
two-wing curve cusped at the nucleus. Atomic units $(a.u.)$ are used, that
is, the Bohr radius $r_{B}$ and the ``Bohr speed'' $v_{B}=2\pi
e^{2}/h=\alpha c$---the electron speed in the ground-state Bohr orbit of the 
$H$ atom---with fine-structure constant $\alpha $ $\approx $ 1/137 and speed
of light $c$. The curve's wing along the positive $z$ axis, $|z|=r$, gives
the \textit{radial} speed, $v_{r}(r)=|v(z)|$, necessary for Sommerfeld's 
\textit{radial} quantization condition,

\begin{equation}
\oint p_{r}dr=m\oint v_{r}(r)dr=n_{r}h.  \tag{3}
\end{equation}
\noindent

\noindent Here $p_{r}$ is the radial momentum, $n_{r}=1,2,...$ is the radial
quantum number, and $h$ is Planck's quantum of action. Integration is over
one period of the radial motion, $z=+\Gamma \rightarrow 0\rightarrow +\Gamma 
$. Graphically, the radial quantization is illustrated in Fig. 3 by the area
under \textit{one} wing of the speed curve. For a line orbit, $l=0$ , the
radial quantum number equals the principal quantum number, $n\equiv
n_{r}+l=n_{r}$.

In order to express the radial quantization in terms of \textit{axial}
motion we employ a ``fold-out factor,'' $\phi =2$, to compensate for the
doubling of integration range in the extension from the radial one-wing
speed curve to the axial two-wing curve. The quantization is thus restated,

\begin{equation}
\frac{1}{\phi }\oint p_{z}dz=\frac{m}{\phi }\oint v(z)dz=n_{z}h,  \tag{3'}
\end{equation}

\noindent with axial quantum number $n_{z}=n_{r}=n$ and integration over the
axial double-wing range, $z=+\Gamma \rightarrow \ -\Gamma \rightarrow \
+\Gamma $. By symmetry we can restrict the axial action integral to one
quarter of the oscillation, say $z=+\Gamma \rightarrow \ 0$,

\begin{equation}
\frac{1}{\phi }\oint p_{z}dz=-\frac{4m}{\phi }\int_{\Gamma }^{0}v(z)dz=-%
\frac{4e}{\phi }\sqrt{2Zm}\int_{\Gamma }^{0}\sqrt{\frac{1}{z}-\frac{1}{
\Gamma }}dz=nh.  \tag{4}
\end{equation}

\noindent Here the electron's motion in the negative $z$ direction is
accounted for by the negative sign. Although the electron's speed through a
point nucleus diverges, $v(0)=\infty $, the action integral, Eq. (4), stays
finite and determines the quantized amplitude $\Gamma _{n}$ of the Coulomb
oscillator. Graphically the amplitude $\Gamma _{n}$ must be such that it
stretches the speed curve horizontally to the extent that the area under one
wing, $A_{n}=nh$, represents the quantized action. The analytic solution,
derived in Appendix A, is

\begin{equation}
\Gamma _{n}=2\frac{r_{B}}{Z}n^{2}.  \tag{5}
\end{equation}

\noindent Inserting Eq. (5) into Eq. (1) yields the quantized energy,

\begin{equation}
E_{n}=-\frac{Z^{2}e^{2}}{\Gamma _{n}}=-\frac{Z^{2}}{n^{2}}R_{y},  \tag{6}
\end{equation}

\noindent in terms of the Rydberg energy unit, $R_{y}=2\pi
^{2}me^{4}/h^{2}=13.6$ $eV$, and in agreement with the energy of the $n$th
Bohr orbit.

Note that Eq. (5) gives the amplitude of the $n$th Coulomb oscillator as 
\textit{twice} the radius of the $n$th Bohr orbit or of the semimajor axis
of an $nl$ Sommerfeld ellipse, $r_{n}=a_{nl}=(r_{B}/Z)n^{2}$. For
comparison, the time-average radial distance of a Kepler orbit\cite{9} of
major and minor semiaxes $a$ and $b$, respectively, is $\langle r\rangle
_{t}=(3a^{2}-b^{2})/(2a)$. A line ellipse ($b=0$) has then $\langle r\rangle
_{t}=\frac{3}{2}a$. Applied to an $nl$ Sommerfeld orbit,\cite{9} its average
size, $\langle r_{nl}\rangle _{t}=(r_{B}/Z)[3n^{2}-l(l+1)]/2$, is in
agreement with the corresponding quantity from quantum mechanics,\cite{10} $%
\langle r_{nl}\rangle =$ $\smallint $ $\psi ^{*}r\psi d^{3}r$. Thus the
time-average radial distance of a Coulomb oscillator is $\langle
r_{n0}\rangle _{t}=\frac{3}{2}(r_{B}/Z)n^{2}$. For the ground state of the
hydrogen atom, $n=1$, this gives $\langle r_{10}\rangle _{t}=\frac{3}{2}
r_{B} $, as is well-known from quantum mechanics.\cite{10}

The concept of the electron's semiclassical Coulomb oscillation is
consistent with the Fermi-contact term of hyperfine interaction for $l=0$
states, which arises from the presence of the electron \textit{inside} the
nucleus. This is familiar from quantum mechanics\cite{11} and can be
interpreted semiclassically as a local-field effect.\cite{12}

To be sure, the extension of Sommerfeld's theory by the Coulomb oscillator
resolves the discrepancy of the old quantum theory with spectroscopy,
mentioned above, only at the \textit{low} end of angular quantum numbers, $%
l=0$. The resolution at the high end---repeal of the circular Bohr orbit, $%
l\neq n$---involves an analysis in terms of space quantization which is
beyond the scope of the present study.\bigskip

\includegraphics[width=5in]{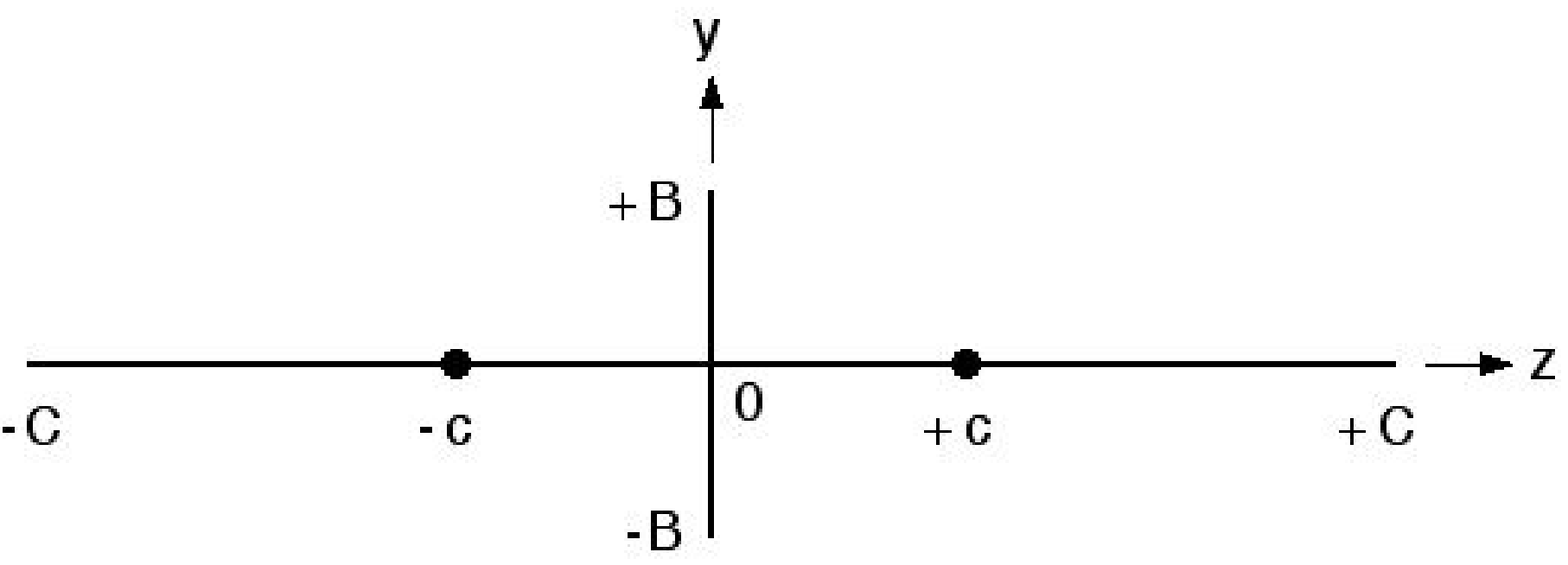}

\begin{quote}
Fig. 4. Axial Coulomb oscillation in an $H_{2}^{+}$ molecule ion between
axial turning points $\pm C$ and through nuclei at $\pm c$; perpendicular
oscillation between lateral turning points $\pm B$ and through midpoint $0$%
.\bigskip
\end{quote}

\section{HYDROGEN MOLECULE ION}

\subsection{Formalism}

A hydrogen molecule ion, $H_{2}^{+}$, consists of two proton nuclei and one
electron. We assume, in adiabatic approximation, the protons located at
fixed positions $z=\pm c$ on the molecular axis (see Fig. 4). The term
``Coulomb oscillator'' denominates again the motion of a point electron, now
along either the line through the protons ($z$ axis) and with turning points 
$\pm C$, or along the perpendicular line through the midpoint ($y$ axis)
with turning points $\pm B$. The molecule ion's total energy $E$ at any
position on the axis, $-C\leqslant z\leqslant +C$, must equal the potential
energy at the turning point $C$,

\begin{equation}
E=\frac{1}{2}mv^{2}-\frac{e^{2}}{|z+c|}-\frac{e^{2}}{|z-c|}+\frac{e^{2}}{2c}
=-\frac{e^{2}}{C+c}-\frac{e^{2}}{C-c}+\frac{e^{2}}{2c}.  \tag{7}
\end{equation}

\noindent For electron positions beyond the protons, $z>c$, this gives an
electron speed

\begin{equation}
v_{out}=\pm \frac{2e}{\sqrt{m}}\sqrt{\frac{z}{z^{2}-c^{2}}-\frac{C}{
C^{2}-c^{2}}}.  \tag{8a}
\end{equation}

\noindent For positions between the protons, $0<z<c$, the corresponding
speed is

\begin{equation}
v_{in}=\pm \frac{2e}{\sqrt{m}}\sqrt{\frac{c}{c^{2}-z^{2}}-\frac{C}{
C^{2}-c^{2}}}.  \tag{8b}
\end{equation}

\noindent The speed expressions will be used in the action integral,

\begin{equation}
\frac{1}{\phi }\oint p_{z}dz=\frac{2m}{\phi }%
\int_{C}^{-C}v(z)dz=A_{z}=A_{out}+A_{in},  \tag{9}
\end{equation}

\noindent with outer contribution

\begin{equation}
A_{out}=\frac{2m}{\phi }\left[
\int_{C}^{c}v_{out}(z)dz+\int_{-c}^{-C}v_{out}(z)dz\right]  \tag{10a}
\end{equation}

\noindent and inner contribution

\begin{equation}
A_{in}=\frac{2m}{\phi }\int_{c}^{-c}v_{in}(z)dz.  \tag{10b}
\end{equation}

\noindent Here $\phi $ is a fold-out factor to be specified below.

If the conditions are such that the electron swings along the $z$ axis
through the midpoint $0$, that is, $v_{in}(0)>0$, then there exists also an
oscillation along the $y$ axis with lateral speed $u$, having the same total
energy,

\begin{equation}
E=\frac{1}{2}mu^{2}-\frac{2e^{2}}{\sqrt{y^{2}+c^{2}}}+\frac{e^{2}}{2c}=-%
\frac{2e^{2}}{\sqrt{B^{2}+c^{2}}}+\frac{e^{2}}{2c}.  \tag{11}
\end{equation}

\noindent Equal energy at the axial and lateral turning points, $E(C)=E(B)$,
Eqs. (7) and (11), determines the latters' geometric dependence,

\begin{equation}
B=\sqrt{C^{2}+\frac{c^{4}}{C^{2}}-3c^{2}}.  \tag{12}
\end{equation}

\noindent Solving Eq. (11) for the lateral speed,

\begin{equation}
u=\pm \frac{2e}{\sqrt{m}}\sqrt{\frac{1}{\sqrt{y^{2}+c^{2}}}-\frac{1}{\sqrt{
B^{2}+c^{2}}},}  \tag{13}
\end{equation}

\noindent provides the integrand of the action integral over a lateral
oscillation,

\begin{equation}
A_{y}=\frac{1}{\phi }\oint p_{y}dy=\frac{2m}{\phi }\int_{B}^{-B}u(y)dy. 
\tag{14}
\end{equation}

Subtraction of the protons' mutual repulsion from the total energy $E$ of
axial or lateral motion, Eqs. (7) and (11), gives the \textit{electronic}
energy,

\begin{equation}
E_{el}=E-\frac{e^{2}}{R},  \tag{15}
\end{equation}

\noindent in its dependence on the proton separation, $R=2c$. This brackets
the molecular problem with known atomic results, Eq. (6), in the limits of $%
R=\infty $ (free $H$ atom, $Z=1$) and $R=0$ (free $He^{+}$ ion, $Z=2$). For
those cases, as well as any proton-proton distance $R$ between, we keep the
action constant,

\begin{equation}
A=A_{z}+A_{y}=nh.  \tag{16}
\end{equation}

\noindent Equation (16) is the \textit{Einstein }quantization condition\cite
{13}---a generalization of Sommerfeld's quantization over separable
variables---where the \textit{quantum sum} equals the sum of action
integrals over topologically independent paths in phase space.\cite{14}%
\bigskip 

\includegraphics[width=5in]{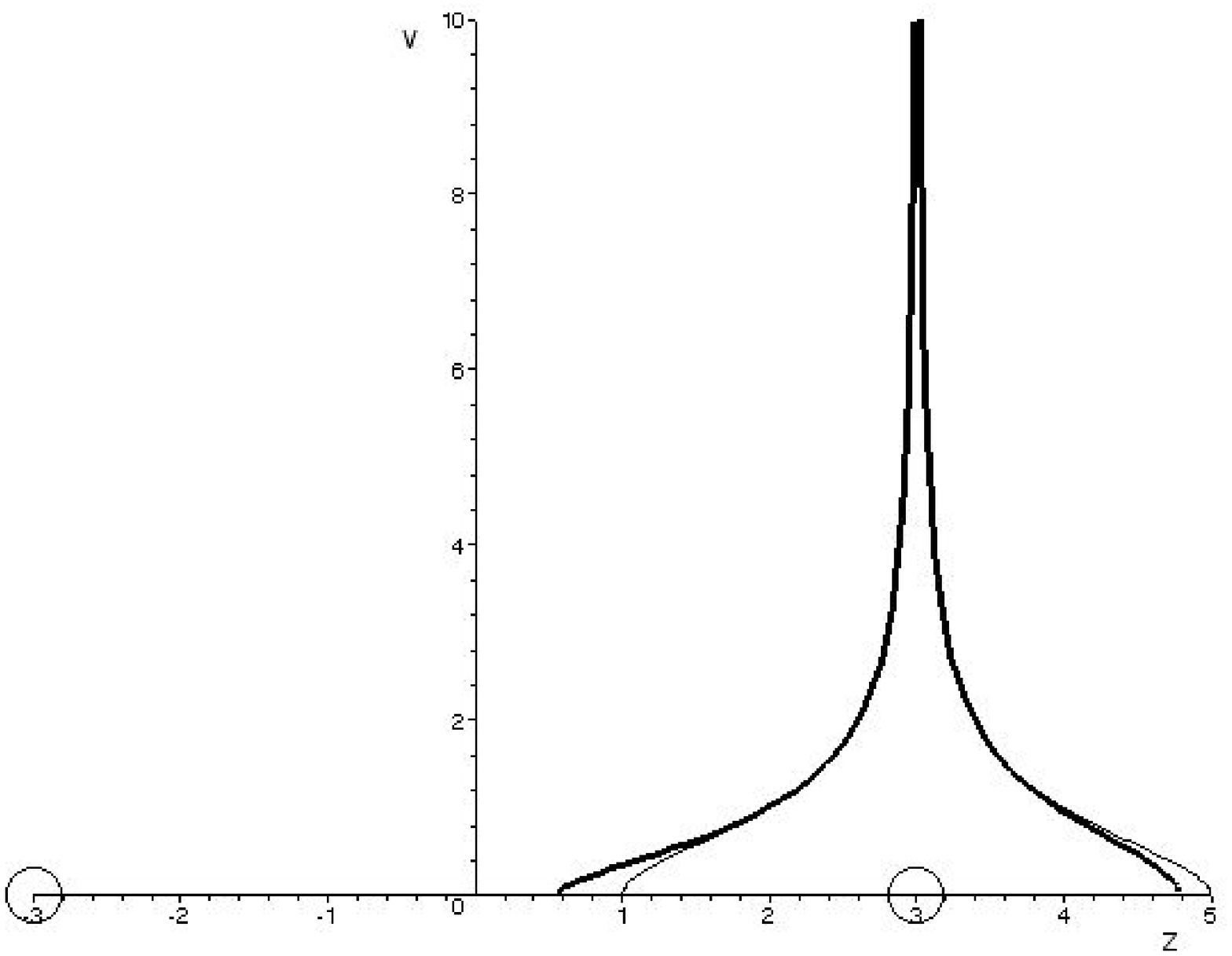}

\begin{quote}
Fig. 5. Axial speed $v$ vs. position $z$ in the ground state of an $H_{2}^{+}
$ molecule ion (bold curve) and, for comparison, of a free $H$ atom (thin
curve). Circles indicate the axial the positions of the nuclei, here with a 
\textit{large} separation, $R=6$ $r_{B}$.\bigskip 
\end{quote}

Here we treat the molecule ion only in its \textit{ground state}, $n=1$.
Analytic solutions of the action integrals, Eqs. (10ab) and (14), are
complicated due to elliptic functions. We therefore integrate numerically
and visualize the integrals by the area under the corresponding speed
curves. The bold curve in Fig. 5 shows the axial electron speed $v(z)$ for a 
\textit{far} proton separation, $R=6$ $r_{B}$. The electron, in its
semiclassical motion, then oscillates only about (and through) the right
proton. The area under the speed curve, Eq. (8ab), proportionally represents
the ground state's unity of action, $m\oint v(z)dz=\phi h$, with a fold-out
factor $\phi =2$ in analogy to the free-atom case, Eq. (4). The thin curve
shows, for comparison, the axial electron speed in a free $H$
atom---familiar from Fig. 3---centered at the same proton position, $+c$.
The pull from the left proton (at $-c$) on the oscillating electron can be
seen by the distortion of the speed function $v(z)$ and the redistribution
of the area under the curve.\bigskip 

\includegraphics[width=5in]{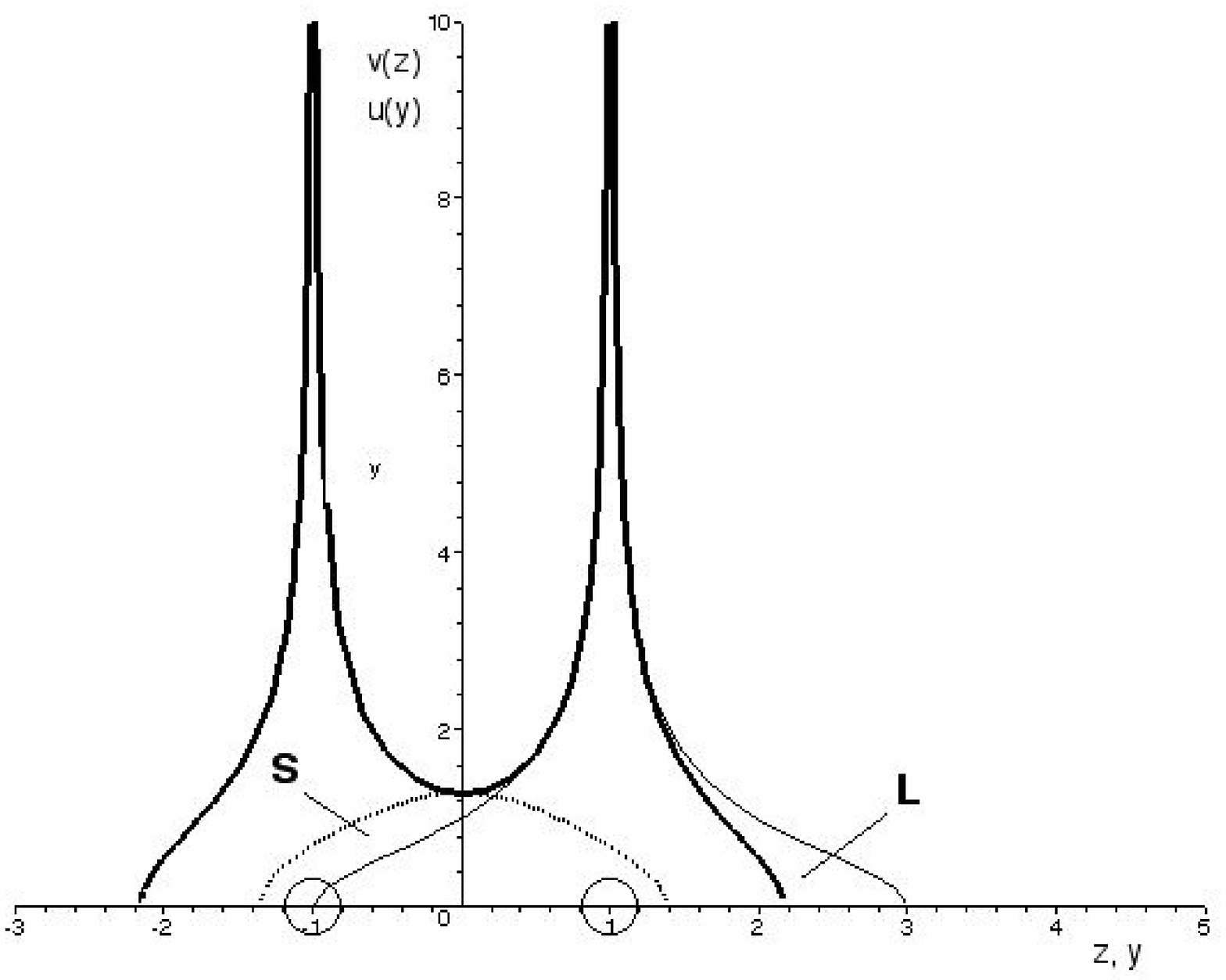}

\begin{quote}
Fig. 6. Axial speed $v$ vs. position $z$ in the ground state of an $H_{2}^{+}
$ molecule ion (M-shaped bold curve) and, for comparison, of a free H atom ($%
\Lambda $-shaped thin curve, centered at the right nucleus). The $\cap $
-shaped dotted curve, centered at $0$, shows the lateral speed $u$ vs. the
perpendicular position $y$ in the molecule. Circles indicate the axial
positions of the nuclei, here with a \textit{small} separation, $R=2$ $r_{B}$%
.\bigskip 
\end{quote}

When, with closer proton separation $R$, as in Fig. 6, the electron swings
past the midpoint $0$, then the single-cusp speed curve $v(z)$ from Fig.
5---akin in shape to letter $\Lambda $---becomes double-cusped (akin to
letter M) and symmetric with respect to the bisector ($y$ axis). Now there
is also a lateral oscillation with speed $u(y)$ having the same total energy 
$E$. The equality of $E$ in both cases can be seen in Fig. 6 by the equality
of axial and lateral speed at the midpoint, $v(0)=u(0)$---a position where
the electron experiences the same potential in either case. For convenience
the lateral speed $u(y)$, though perpendicular to the proton axis, is
displayed in Fig. 6 together with the axial speed $v(z)$. The lateral speed
curve $u(y),$ drawn dotted, is readily recognized by its dome shape ($\cap $
). At the bifurcation value of the proton separation, $\check{R}$, where the
electron starts swinging though the midpoint $0$, the axial speed curve $%
v(z) $ changes from its one-centered $\Lambda $ shape to a two-centered M
shape. The \textit{area} under the axial speed curve then abruptly doubles, M%
$(\check{R}-\delta )$ $\approx 2\Lambda (\check{R}+\delta )$, upon a very
small change in proton separation, $\delta \ll $ $\check{R}$. In order to
keep the action integral continuous at $\check{R}$, the sudden area doubling
is compensated by a corresponding doubling of the fold-out factor from $\phi 
$ $=2$ for $R$ $>$ $\check{R}$ to $\phi =4$ for $R$ $<\check{R}.$ The same
fold-out factor, $\phi $ $=4$, must be used for the lateral action integral $%
A_{y}$, Eq. (14), as will become clear shortly.

The unity of action, $A=1h$, is visualized again in Fig. 6. To this end we
compare the right half of the bold M-shape curve of axial electron speed in
the molecule with the thin curve $\Lambda (H)$ of the axial speed in a free $%
H\,$atom positioned at the right nucleus, $+c$. Due to attraction from the
left nucleus, the right wing of the M curve is smaller than that of $\Lambda
(H)$ by the area of lobe $L$. On the other hand, the left flank of the $%
\Lambda (H)$ curve that extends over the negative $z$ axis is smaller than
the left wing of the lateral speed curve $\cap $ by the area of slice $S$.
The area under the free-atom curve is then 
\begin{equation}
\Lambda (H)\approx \frac{1}{2}\text{M}+L+\frac{1}{2}\cap -S.  \tag{17a}
\end{equation}

\noindent The areas of lobe and slice are comparable, 
\begin{equation}
L\approx S.  \tag{17b}
\end{equation}

\noindent When the tiny notch to the right of the saddle point of M is taken
into account, then the approximations (17ab) become equations and combine to 
\begin{equation}
\text{M}+\cap =2\Lambda (H).  \tag{18}
\end{equation}

\noindent The area under both the axial and lateral speed curves is thus
four times the area under \textit{one} wing of the free-atom curve, M$+\cap
=4\times \frac{1}{2}\Lambda (H)$. Since the latter represents one quantum of
action, $h$, the combined area M$+\cap $ visualizes its \textit{double}
fold-out, $\phi =4$.

With very close proximity of the nuclei, $R\rightarrow 0$, the crests of the
M curve start merging while its saddle point, $v(0)$, keeps rising. In the $%
R=0$ limit of fusing nuclei the axial electron speed becomes that of a free $%
He^{+}$ ion, M$\rightarrow $\ $\Lambda (He^{+})$, familiar from Fig. 3.
Concurrently, the lateral speed curve $\cap $ rises at its peak, $u(0)$ $=$ $%
v(0)$, and narrows at its base until it, too, turns into the speed curve of
the free $He^{+}$ ion, $\cap \rightarrow $\ $\Lambda (He^{+})$. In the $R=0$
limit the three curves merge, M$(0)=\cap (0)$ $=$ $\Lambda (He^{+})$.

The results of the Coulomb-oscillator approach will be compared with another
semiclassical calculation of $H_{2}^{+}$, by Strand and Reinhardt.\cite{15}
These authors, like Pauli,\cite{4} separate the equation of motion in
spheroidal coordinates, $\xi $ $=(r_{+}+r_{-})/2c$ , $\eta =(r_{+}-r_{-})/2c$
and $\varphi $ by virtue of the constants of the motion: total energy $E$,
angular momentum $\mathbf{M}$ about the $z$ axis, and a component of the
bifocal Runge-Lenz vector, $\Omega _{c}$.\cite{16} Here $r_{+}$ $(r_{-})$ is
the distance of the electron from the nucleus at $+c$ $(-c).$ Strand and
Reinhardt (SR) solve the ensuing one-dimensional differential equations with
classical Poisson-bracket techniques. They find the electron's trajectories
conditionally periodic\cite{17} and regionally confined due to restrictions
from $E$, $\mathbf{M}$ and $\Omega _{c}$. However, unlike Pauli, who used
Sommerfeld quantization, SR employ the Einstein-Brillouin-Keller (EBK)
quantization conditions,

\begin{equation}
A_{j}=\oint p_{j}dj=(n_{j}+\frac{1}{2})h,\text{ }j=\xi ,\text{ }\eta 
\tag{19ab}
\end{equation}

\noindent and

\begin{equation}
A_{\varphi }=\oint p_{\varphi }d\varphi =n_{\varphi }h.  \tag{19c}
\end{equation}

The background of EBK quantization touches on the foundations of classical
and quantum mechanics.\cite{18} For the present purpose its essential
rationale may be summarized as follows: A semiclassical treatment envolves
turning points of radial, or other librating motion. Any tunneling through
``forbidden'' regions of negative kinetic energy is ruled out. Viewed in
terms of the quantum mechanical WKB (Wentzel-Kramers-Brillouin)
approximation, the hard reflection of a wavefunction at a turning point
corresponds to a so-called ``loss'' of phase (phase shift by $\pi $)
compared to the soft reflection caused by tunneling (phase shift by $\pi /2$
). This shortcoming can be remedied with EBK quantization conditions by
addition of a value of 1/4, for each librational turning point, to the
corresponding quantum number. Such is the case for the electron's elliptical
and hyperbolic librations in the above quantization, Eq. (19ab), but not for
a rotation about the $z$ axis, Eq. (19c). Strand and Reinhardt call these
quantization conditions ``primitive'' to distinguish them from more
sophisticated ones, specified below.

A semiclassical treatment of a free $H$ atom with EBK quantization has
recently been presented in these pages.\cite{19} In this case the isotropic
symmetry permits a separation of variables in spherical coordinates, $r$, $%
\theta $ and $\varphi $, and the EBK quantization conditions are like Eqs.
(19abc) except for $j=r,$ $\theta $. The atomic ground state is
characterized by the quantum numbers ($n_{r},$ $n_{\theta },$ $n_{\varphi
})=(0,0,0)$. Accordingly, the action in the atom's ground state, $A_{1}=$ $1h
$, is attributed only to tunneling at the radial and latitudinal turning
points (the former being the nucleus). Applying EBK quantization to the
ground state of $H_{2}^{+}$, denoted $1s\Sigma _{g}$ in molecular
spectroscopy, SR likewise assign the quantum numbers $(n_{\xi },n_{\eta
},n_{\varphi })=(0,0,0)$.

\subsection{Results}

Energies of the Coulomb oscillator (CO), in adiabatic dependence on the
proton separation, are listed in Appendix B and shown in Fig. 7 in
comparison with exact quantum mechanical (QM) results and the semiclassical
calculation by SR.\cite{15} The lower part of Fig. 7 shows the electronic
energy $E_{el}(R)$ of the ground state, $1s\Sigma _{g}$. At \textit{large}
proton separations, $R>6$ $r_{B}$, both the SR calculation (circles) and the
CO approach (crosses) agree excellently with the exact QM values (curve).
This is the situation where the electron stays near one nucleus (see Fig.
5). As Fig. 7 further shows, such agreement ceases once the classical
electron motion leads beyond the molecular bisector, which happens for
proton separations below the bifurcation value, $R<\check{R}\approx 5.57$ $%
r_{B}$ (see Fig. 6). Interestingly, the deviations of CO and SR from QM are
opposite over the entire range. The SR results are \textit{discontinuous} at
a certain proton separation, $R^{*}\approx 1.38$ $r_{B}$. Remarkably, at (or
near) that value, $R^{*}$, the CO energy crosses the curve of the QM
solution.
\bigskip 

\includegraphics[width=5in]{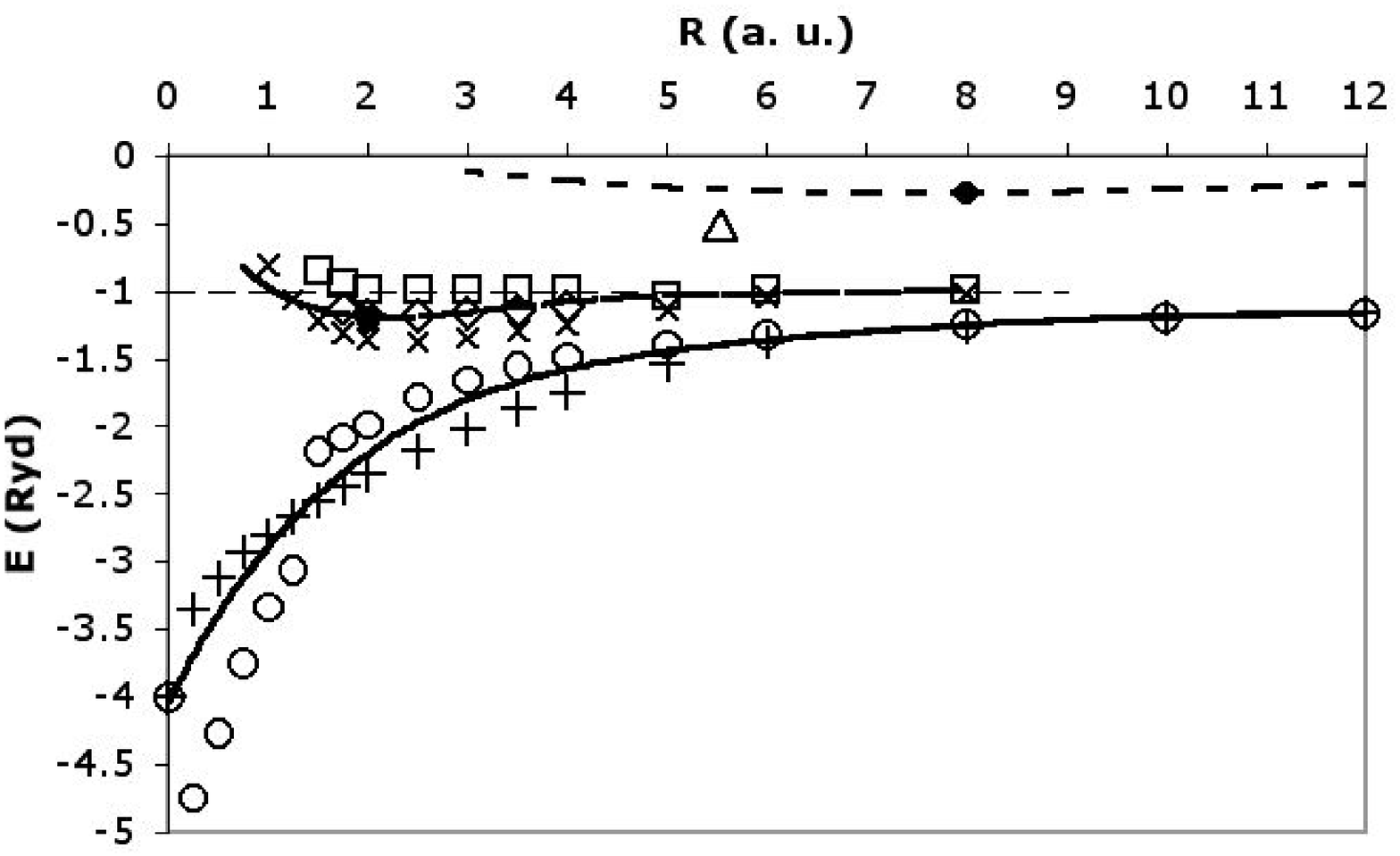}

\begin{quote}
Fig. 7. Dependence of energies of $H_{2}^{+}$ on internuclear distance $R$.

\textit{Bottom:} Electronic energy $E_{el}$ from quantum mechanics (QM,
solid curve), primitive semiclassical quantization by Strand and Reinhardt
(SR, $\bigcirc $) and the present Coulomb-oscillator approach (CO, +).

\textit{Middle}: Total energy $E$ of the $1s\Sigma _{g}$ ground state from
QM (dashed curve) with minimum ($\bullet $), values by SR ($\square $) and
CO ($\times $) and their average ($\diamond $), energy of a free $H$ atom
(dashed horizontal line).

\textit{Top}: Total energy $E$ of the $2p\Pi _{u}$ state from QM (dotted
curve) with minimum ($\bullet $), and historical value by Pauli and Niessen (%
$\bigtriangleup $).\bigskip 
\end{quote}

Adding to the electronic energy $E_{el}$ the proton-proton repulsion gives
the \textit{total} energy $E$, Eq. (15). The middle part of Fig. 7 shows by
the dashed curve the exact total energy $E(R)$ of the $1s\Sigma _{g}$ ground
state, obtained from QM and, by symbols, the corresponding CO and SR values.
The solid dot at the minimum of the curve shows the QM equilibrium energy,
in agreement with experiment, $E_{0}=-1.20$ $R_{y}$, and the equilibrium
internuclear distance, $R_{0}=2.00$ $r_{B}$. The CO energy ($\times $) comes
out too low, due to the inaccuracy of its $E_{el}$, with an equilibrium
value $E_{0}(CO)$ $\approx -1.38$ $R_{y}$ at $R_{0}(CO)$ $\approx $ $2.5$ $%
r_{B}$. Conversely, the SR energy ($\square $) comes out too high with $%
E_{0}(SR)$ $\approx -1.05$ $R_{y}$ at $R_{0}(SR)$ $\approx $ $5$ $r_{B}$.
Since the CO and SR results deviate about equal and oppositely from QM,
their \textit{average} ($\Diamond $) is close to the exact values with a
minimum of $E_{0}[\frac{1}{2}(CO+SR)]$ $\approx -1.19$ $R_{y}$ at $R_{0}[%
\frac{1}{2}(CO+SR)]$ $\approx $ $2.5$ $r_{B}$.

The molecular binding energy is the difference of $E$ in the molecule and in
the constituting atoms, here, $E_{0}(H_{2}^{+})-$ $E(H)$. The energy of a
free hydrogen atom, $E(H)=-1$ $R_{y}$, is indicated in Fig. 7 by the fine
horizontal line. Both semiclassical treatments, CO and (barely) SR, yield
molecular binding energies with \textit{negative} values and thus a \textit{%
\ \ stable} molecule ion. Why are they more successful than the early
attempts, in the 1920s, by Pauli,\cite{4} and independently Niessen,\cite{20}
with the Sommerfeld quantization conditions of the old quantum theory?

For the same reason that Sommerfeld had excluded the angular quantum number $%
l=0$ for the $H$ atom---avoidance of electron collision with the
nucleus---both Pauli and Niessen excluded electron motion in the nuclear
plane of the $H_{2}^{+}$ molecule ion. They then found the lowest admissible
quantum state to be $(n_{\xi },$ $n_{\eta },$ $n_{\varphi })=(0,1,1)$,
denoted $2p\Pi _{u}$ in molecular spectroscopy, with $E_{0}(P,N)=-0.52$ $%
R_{y}$ at $R_{0}(P,N)=5.53$ $\pm 0.01$ $r_{B}$, depicted by the triangle in
the top part of Fig. 7. The QM energy\cite{21} of that quantum state is $%
E_{0}(2p\Pi _{u})=-0.27$ $R_{y}$ at $R_{0}(2p\Pi _{u})$ $\approx $ $8$ $%
r_{B},$ marked by the solid dot at the minimum of the dotted curve. Since
both these energy values are higher than the ground state of a free hydrogen
atom, they give rise to \textit{positive} molecular binding energies and
thus to spontaneous dissociation, $H_{2}^{+}(2p\Pi _{u})$ $\rightarrow $ $H$ 
$+$ $H^{+}$. Qualitatively, Pauli's and Niessen's finding of energetic 
\textit{instability} is borne out by quantum mechanics for this exited state
of $H_{2}^{+}$ (Pauli's argument\cite{4} about ``dynamical stability''
notwithstanding). The deviation of their historical value $(\triangle )$
from the (dotted) QM curve is remarkably small---comparable to those of the
CO and SR results for the ground state. Pauli's and Niessen's misfortune,
though, was that they \textit{\ misinterpreted} their result as the molecule
ion's ground state---an assessment with fateful consequences in the
development of quantum theory.

\subsection{Discussion}

Why do the semiclassical results of the Coulomb oscillator and of SR's
quantization deviate from the QM solution of the $H_{2}^{+}$ molecule ion?
Strand and Reinhardt explain the deviation of their primitive quantization
from QM with effective potential barriers arising from constrictions due to
conservation of both the energy $E$ and the bifocal Runge-Lenz component $%
\Omega _{c}$ (the angular momentum vanishes for the ground state, $\mathbf{M}%
=0$). The most drastic consequence of those barriers is the discontinuity of 
$E_{el}$ at $R^{*}$ (see Fig. 7) and the large deviations at closer proton
separation, $R<R^{*}$. While the simulation of quantum mechanical tunneling
beyond the semiclassical turning points of librations is adequately achieved
by EKB quantization under the isotropic symmetry of a free $H$ \textit{atom},%
\cite{19} Eq. (19) is less successful under the lower symmetry of $H_{2}^{+}$
. When SR remedy the situation with ``unified'' semiclassical quantization
conditions, then $E_{el}$ agrees, for all practical purposes, with QM. Those
unified quantization condition, going well beyond Eq. (19), are
sophisticated in their dependence on $E_{el}$, $\Omega _{c},$ and the
hyperbolic turning points $\eta _{\pm }$. They will not be discussed
here.

The reason for the deviation of the CO results from the QM values is the 
\textit{neclect} of the electron's \textit{wave} nature in the underlying
quantization condition, Eq. (16). In proposing his wave hypothesis de Broglie%
\cite{22} already showed that the quantization condition of the Bohr model, $%
A_{n}=nh$, is equivalent to $n$ standing waves along the $nth$ Bohr orbit.
If $s$ denotes the position along the Bohr orbit, then the de Broglie wave
can be expressed as $w(s)=\sin [2\pi a_{n}(s)/h]$, with the variable $%
a_{n}(s)=(A_{n}/S_{n})\int_{0}^{s}ds^{\prime }$ along the orbit's
circumference $S_{n}=2\pi r_{n}$. A generalization gives the de
Broglie wave of the Coulomb
oscillator of a free $H$ \textit{atom} in the ground state ($n=1$), 
\begin{equation}
w(z)=\cos [2\pi a(z)/h]  \tag{20a}
\end{equation}

\noindent with 
\begin{equation}
a(z)=\frac{m}{\phi }\int_{0}^{z}v(z^{\prime })dz^{\prime }  \tag{20b}
\end{equation}
where $v(z^{\prime })$ is the speed from Eq. (2) and $\phi$ the fold-out factor from
Eqs. (4), (9) and (14).
\bigskip 

\includegraphics[width=5in]{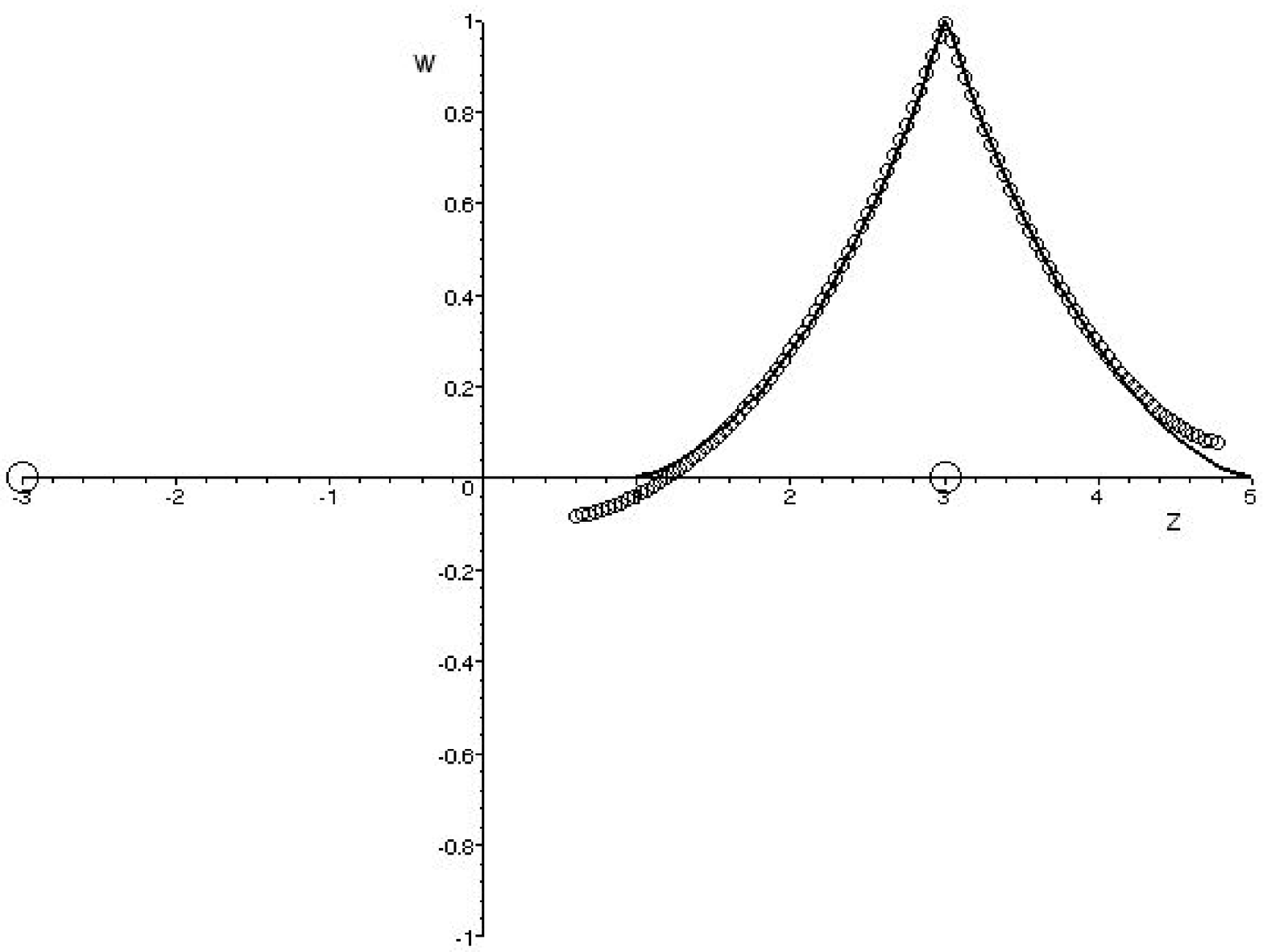}

\begin{quote}
Fig. 8. de Broglie wave of the free $H$ atom Coulomb oscillator (curve) and
of the $H_{2}^{+}$ Coulomb oscillator (small circles) for the same proton
separation as in Fig. 5, $R=6$ $r_{B}$. Large circles indicate the axial
positions of the nuclei.\bigskip 
\end{quote}

This de Broglie wave, shown by the line curve in Fig. 8 for a free $H$ atom 
positioned at $+c$, has peak at the nucleus---like the QM radial wave function---and 
a node at each turning point. By the above characterization,
those turning points are ``soft,'' and it is their softness that ensures the
exact energy of the free atom. However, when Eq. (20ab) is applied to $%
H_{2}^{+}$ for proton separations beyond the bifurcation value, $R>\check{R}$
, with axial speed from Eq. (8ab) and integration away from the occupied
nucleus, $\int_{c}^{z}...$, then the de Broglie wave is found to be
``truncated'' (no nodes) at both the outer and inner turning point (see Fig.
8, small circles). Those turning points are ``hard'' and give rise to
incorrect energies. Qualitatively, an augmentation of the truncated de
Broglie wave with (exponential) ``tunneling tails,'' determined by the
negative kinetic energy in the classically ``forbidden'' region, would
``soften'' the turning points. This would give rise to an \textit{effective}
far turning point farther out, $C_{eff}$ $>C$, and accordingly raise the CO
energy, Eq. (7), toward the QM result.\bigskip 

\includegraphics[width=5in]{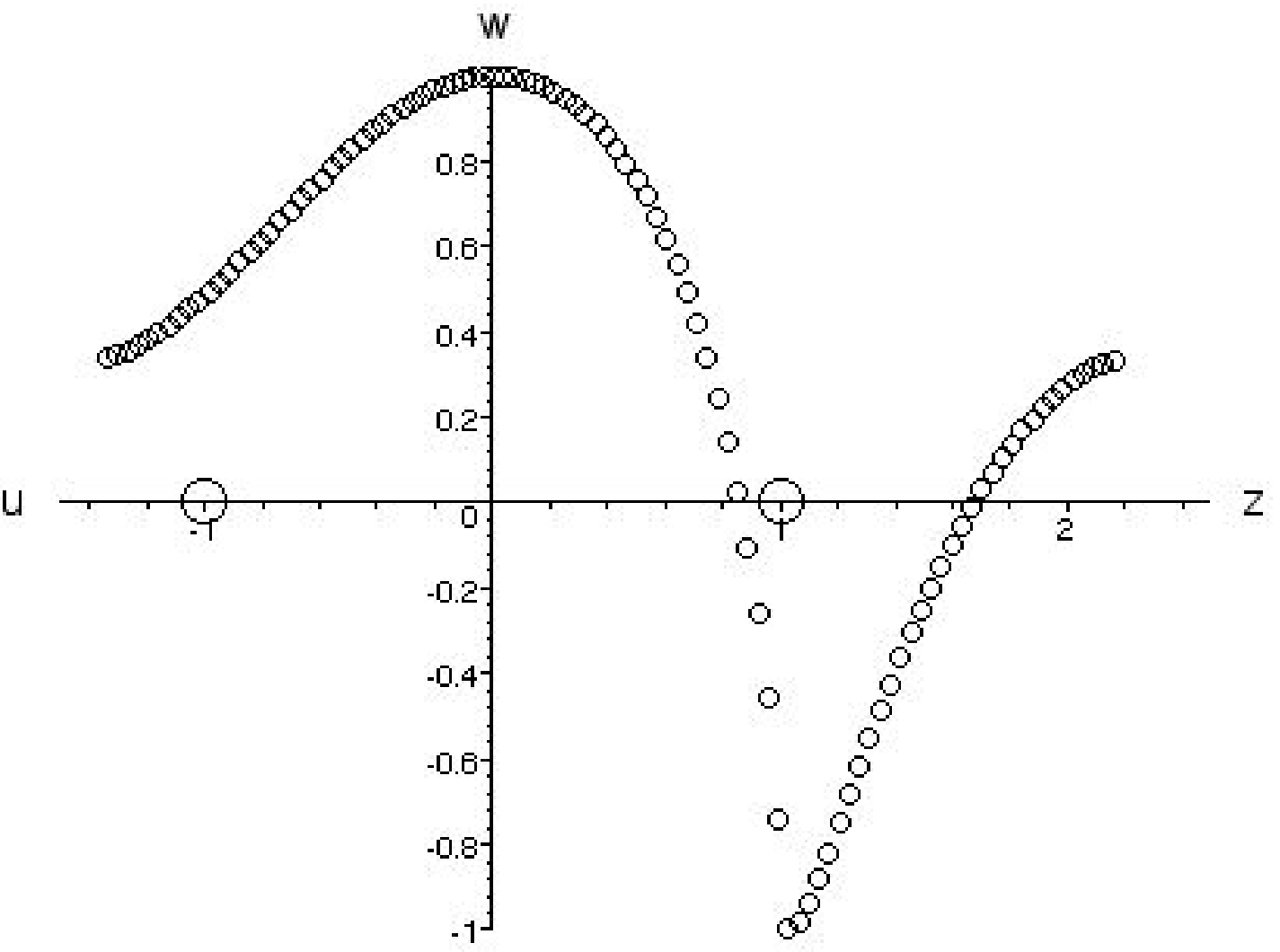}

\begin{quote}
Fig. 9. de Broglie wave of the $H_{2}^{+}$ Coulomb oscillator along axial
distance $OC$ from Fig. 4 (right side of graph) and lateral distance $OB$
(left side). Large circles indicate the axial positions of the nuclei, here
with the same separation as in Fig. 6, $R=2$ $r_{B}$.\bigskip 
\end{quote}

When the proton separation is below bifurcation, $R<\check{R}$, then Eq.
(20b) should be integrated from the midpoint $0$ in both the axial and
lateral direction rather than favoring one nucleus with the crest of the de
Broglie wave. Again, the de Broglie wave is found to be truncated at the
axial and lateral turning points, $C$ and $B$, respectively (see Fig. 9). An
exception exits for the proton separation $R^{*}$. The de Broglie wave,
shown in Fig. 10, then has a minimum at both turning points, $w(C)=w(B)=-1$,
according to action values of $A_{z}=\frac{3}{4}h$ and $A_{y}=\frac{1}{4}h$.
Such turning points seem to be ``benign''---reminiscent of the soft turning
points in the free-atom case---and cause the CO energy $E_{el}(R^{*})$ in
Fig. 7 to be \textit{exact}. At still smaller proton separation, $R<R^{*}$,
the de Broglie wave is truncated again (not shown). The limit $R=0$
corresponds to a Coulomb oscillator in the free $He^{+}$ ion which, like in
the free $H$ atom for $R=\infty $, has de Broglie nodes at the turning
points and an exact energy value.\bigskip 

\includegraphics[width=5in]{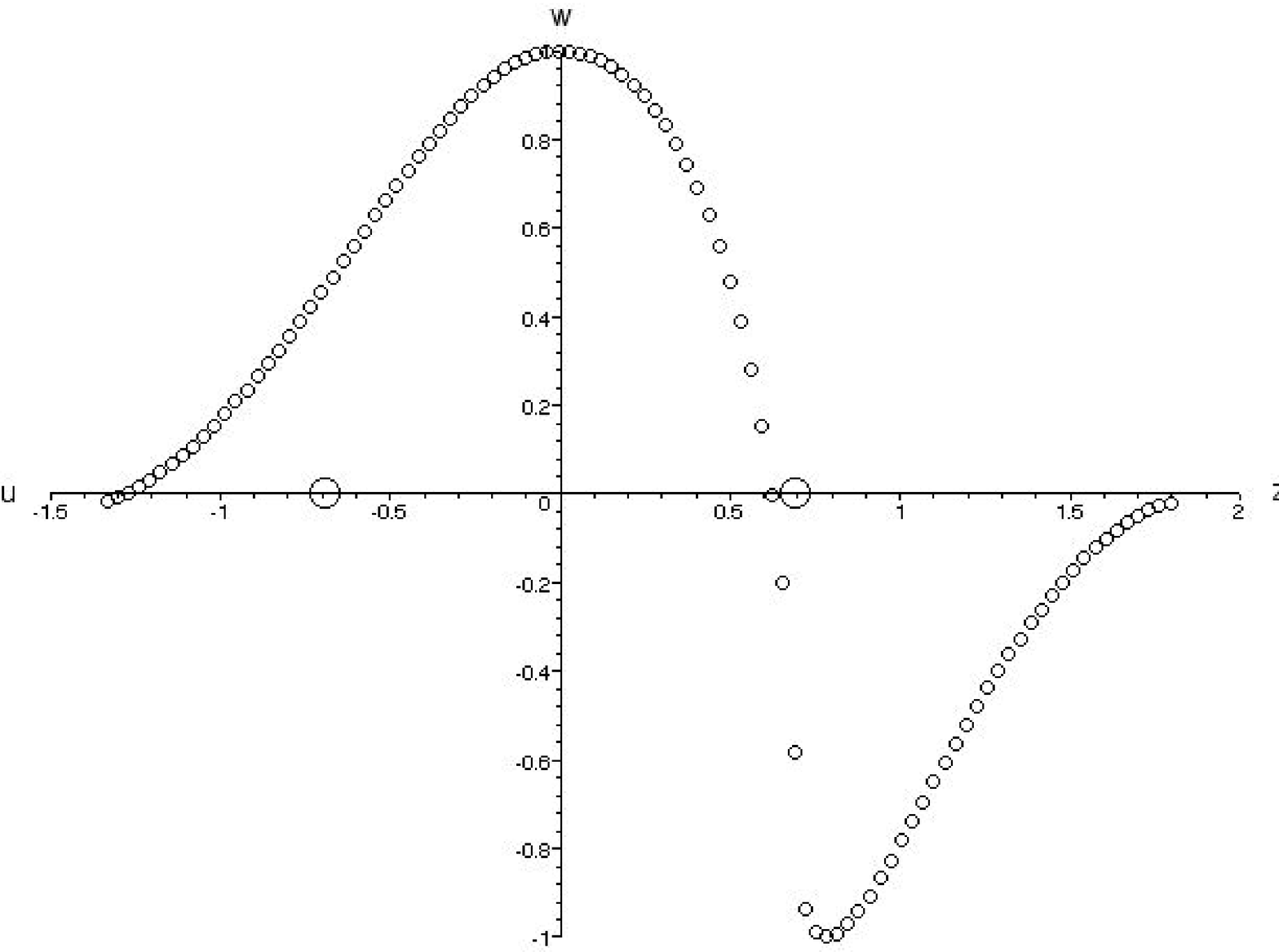}

\begin{quote}
Fig. 10. Same as Fig. 9 but for the proton separation $R^{*}=1.38$ $r_{B}$
where the semiclasssical energy is exact.\bigskip 
\end{quote}

If the explanation that the CO energy $E_{el}(R)$ deviates from the QM curve
because of the \textit{neglect} of \textit{wave} effects is valid, then this
sheds new light on the SR results. The opposite sign of the CO and SR
deviations then suggests that SR's primitive EBK quantization, while
appropriate for a free atom, simulates \textit{too much} \textit{wave effects%
} under the lower symmetry of the $H_{2}^{+}$ molecule ion. However, the 
\textit{average} of both those semiclassical quantizations seems to be an
excellent compromise, as evidenced by the close agreement of the
corresponding total energy $(\Diamond )$ with the (dashed) QM curve in Fig.
7.

In conclusion, semiclassical quantization can rise to Gutzwiller's challenge
and ``explain'' the chemical bond in the paradigm molecule,\cite{23} $
H_{2}^{+}$, by a combination of classical mechanics, quantization, and
moderate wave effects.

\section{\noindent \textbf{ACKNOWLEDGMENTS}}

I thank Duane Siemens and Ernst Mohler for valuable discussions. Many thanks
to Preston Jones for help with computer integration and graphics. I also
thank Professor Gutzwiller for advice and his kind encouragement.

\section{APPENDIX A: QUANTIZATION}

By Eq. (4) the action integral of the atomic Coulomb oscillator is

\begin{equation}
A=\frac{1}{\phi }\oint p_{z}dz=-\frac{4e}{\phi }\sqrt{2Zm}\int_{\Gamma }^{0}%
\sqrt{\frac{1}{z}-\frac{1}{\Gamma }}dz.  \tag{4'}
\end{equation}

\noindent For a comparison with integral tables we change notation to $x=z$
and use the abbreviation $a=-1/\Gamma $. Then

\begin{equation}
\int \sqrt{\frac{1}{z}-\frac{1}{\Gamma }}dz=\int \sqrt{\frac{1}{x}+a}dx=\int 
\frac{\sqrt{X}}{x}dx  \tag{21}
\end{equation}

\noindent with $X=ax^{2}+x$. Integration tables give

\begin{equation}
\int \frac{\sqrt{X}}{x}dx=\sqrt{X}+\frac{1}{2}\int \frac{dx}{\sqrt{X}}. 
\tag{22}
\end{equation}

\noindent The first term on the rhs, evaluated at the limits of the $%
\int_{\Gamma }^{0}$ integration, vanishes. The last integral in Eq. (22),
tabulated as

\begin{equation}
\int \frac{dx}{\sqrt{X}}=(-\sqrt{\Gamma })\arcsin (1-2x/\Gamma ),  \tag{23}
\end{equation}

\noindent and evaluated at the limits, $x=0$ and $x=\Gamma $, contributes

\begin{equation}
-[\arcsin (1)-\arcsin (-1)]\sqrt{\Gamma }=-\pi \sqrt{\Gamma }.  \tag{24}
\end{equation}

\noindent Combining Eqs. (4'), (22) and (24), together with a fold-out
factor $\phi $ = 2, gives the action integral, to be equated with the
Sommerfeld quantization condition,

\begin{equation}
A_{n}=-\frac{4e}{\phi }\sqrt{2Zm}\frac{1}{2}(-\pi \sqrt{\Gamma })=nh. 
\tag{25}
\end{equation}

\noindent We square Eq. (25) and solve for the amplitude of the quantized
Coulomb oscillator,

\begin{equation}
\Gamma _{n}=2\frac{r_{B}}{Z}n^{2},  \tag{26}
\end{equation}

\noindent in terms of the Bohr radius $r_{B}$.

\section{APPENDIX B: DATA}

\noindent TABLE I. Electronic ground-state energy $E_{el}$ for various
nuclear separations $R$ of the $H_{2}^{+}$ molecule from quantum-mechanical
calculations (QM, Ref 21), the ``primitive'' semiclassical quantization of
Strand and Reinhardt (SR, Ref. 15), and the present Coulomb-oscillator
approach (CO).

\begin{tabular}{llll}
R $(a.u.)$ & QM $(R_{y})$ & SR $(R_{y})$ & CO $(R_{y})$ \\ 
0.00 & -4.00 & -4.00 & -4.00 \\ 
0.25 & -3.80 & -4.75 & -3.35 \\ 
0.50 & -3.47 & -4.27 & -3.12 \\ 
0.75 & -3.11 & -3.76 & -2.93 \\ 
1.00 & -2.90 & -3.34 & -2.81 \\ 
1.25 & -2.68 & -3.07 & -2.67 \\ 
1.50 & -2.50 & -2.19 & -2.55 \\ 
1.75 & -2.34 & -2.09 & -2.45 \\ 
2.00 & -2.21 & -1.99 & -2.36 \\ 
2.50 & -1.99 & -1.79 & -2.18 \\ 
3.00 & -1.82 & -1.66 & -2.02 \\ 
3.50 & -1.69 & -1.56 & -1.87 \\ 
4.00 & -1.59 & -1.49 & -1.75 \\ 
5.00 & -1.45 & -1.40 & -1.54 \\ 
6.00 & -1.36 & -1.33 & -1.38 \\ 
8.00 & -1.26 & -1.25 & -1.27 \\ 
10.00 & -1.20 & -1.20 & -1.20 \\ 
12.00 & -1.17 & -1.17 & -1.17
\end{tabular}


\begin{thebibliography}{99}
\bibitem{1}  Other failures were its inability to give the brightness of
spectral lines and the unsuccessful extension to the $He$ atom.
\bibitem{3}  Instead of the traditional notation $k$ for the angular quantum
number in the old quantum theory, we will use the letter $l$ $(=k)$ to
facilitate the connection with quantum mechanics.
\bibitem{7}  This length of semiminor axis has been modified from
Sommerfeld's original expression, $b_{nl}=(r_{B}/Z)nk$, to achieve agreement
with expressions from quantum mechanics.
\bibitem{8}  If the nuclear charge is uniformly distributed, then $Z^{\prime
}(r)=(r/r_{N})^{3}Z$, and the electron's motion inside the nucleus is
harmonic.
\bibitem{14}  Eq. (3), introduced as Sommerfeld quantization for
familiarity's sake, is also an Einstein quantization condition.
\bibitem{17}  The trajectory will eventually revisit an arbitrarily close
neighborhood of any previous point.
\end{thebibliography}
\end{document}